\documentclass[a4paper]{jpconf}
\usepackage{graphicx, amsmath, amsfonts,amssymb}
\usepackage[square,sort&compress,numbers]{natbib}
\begin{document}
\newcommand{\ket}[1]{| {#1}\rangle}
\newcommand{\bra}[1]{\langle {#1}|}
\newcommand{\bvec}[1]{\mathbf{#1}}
\title{A Numerical Study of the Superconducting Proximity Effect in Topological Surface States}

\author{Roland Grein$^{1,2}$, Jens Michelsen$^1$, Matthias Eschrig$^2$}

\address{$^1$ Institut f\"ur Theoretische Festk\"orperphysik
and DFG-Center for Functional Nanostructures,
Karlsruhe Institute of Technology, D-76128 Karlsruhe, Germany}
\address{$^2$ SEPnet and Hubbard Theory Consortium, Department of Physics, Royal Holloway, University of London, Egham, Surrey TW20 0EX, United Kingdom}

\ead{roland.grein@kit.edu}

\begin{abstract}
We study the superconducting proximity effect induced in the surface states of the 3-d topological insulator Bi$_2$Se$_3$ by a singlet, s-wave superconductor deposited on its surface. To this effect, the $\mathbf{k}\cdot\mathbf{p}$-Hamiltonian of Bi$_2$Se$_3$ and the BCS-Hamiltonian are mapped onto tight-binding chains which we couple through a transfer-Hamiltonian at the interface. We then employ the Recursive Green's Function technique to obtain the local spectral function and infer the dispersion of the interface-states from it. In agreement with earlier microscopic studies of this problem, we find that the Fu-Kane model is a reasonable approximation at energies $\epsilon\ll \Delta_{\rm SC}$. However, for energies close to the SC bulk gap, the Fu-Kane model is expected to break down. Indeed, our numerical calculations show strong modifications of the interface-state dispersion for $\epsilon \gtrsim \Delta_{\rm SC} $. We find that the proximity effect can be strong enough to induce a gap in the surface state that is comparable to the superconducting gap. An analysis of the spatial profile of the states shows that their weight shifts towards the SC as the coupling strength increases. We conclude that an intermediate coupling is ideal for realising the Fu-Kane scenario.\end{abstract}
\section{Introduction}
The research effort devoted to topological insulators has grown rapidly since their initial discovery in 2007 \cite{Koenig:2007hc} and is presently developing in a multitude of directions \cite{Hasan:2010fk, RevModPhys.83.1057}, one of them being the superconducting proximity effect. It was argued that in superconducting heterostructures involving 3-dimensional topological insulators, the physics of the interface region can effectively be described in terms of a topological superconductor, with Majorana-Fermions at line-junctions and vortex-cores \cite{Fu:2008ly}. These excitations obey non-abelian statistics and are thus highly relevant to the field of topological quantum computation \cite{Nayak:2008tw}. Apart from proximity structures, topological superconductivity can also be realised in non-centrosymmetric materials, which feature spin-current carrying surface states \cite{Vorontsov:2008jl}. Such surface states are characteristic for topological materials, where the electronic structure is characterised by non-trivial topological invariants.

In their initial work, Fu and Kane described the proximity effect in a topological insulator by simply including a singlet, s-wave pair-potential in the effective Hamiltonian of the 2-dimensional surface state \cite{Fu:2008ly}:
\begin{align} \mathcal{H}_{\rm surface}=\left(\begin{array}{cc}H_0(\bvec{k}) & \Delta i\sigma_2 \\ -\Delta i\sigma_2 & -H_0^*(-\bvec{k}) \end{array}\right),\quad H_0(\bvec{k})=v_{\rm F}( k_x \sigma_1+k_y\sigma_2)-\mu.  \end{align}
$v_{\rm F}$, $\Delta$ and $\mu$ can be considered as free parameters that we will use to fit the dispersion of this model to our numerical results.

While the proximity effect is known to induce pairing correlations which suppress the density of states and, under certain circumstances, opens up a gap in the quasiparticle spectrum of an adjacent material, it does not induce a pair potential as such \cite{RevModPhys.36.225}. The latter would require a finite pairing interaction on the normal side, as can easily be seen from the gap-equation.
Thus, there was no obvious answer to the question whether this model can be justified from a microscopic theory of the proximity effect, and a first investigation was provided by Stanescu et al. \cite{Stanescu:2010gb}. Their work was based on a tight-binding model describing the whole structure within a transfer-Hamiltonian approach. The authors concluded that the Fu-Kane model is a valid low-energy theory and that the inclusion of a pair-potential in the surface state Hamiltonian can be derived from integrating out the superconducting part of the structure and including it as a self-energy within a Green's function formalism. Since the relevant states are localised at the interface, this self-energy formally appears as an effective order parameter in the surface state Hamiltonian \cite{Sau:2010cq}. A similiar study has also been performed for the edge-states of 2D TIs \cite{Black-Schaffer:2011fk}.

A different approach was presented a little later by Lababidi et al. \cite{Lababidi:2011fk}, using a numerical method to solve the Bogoliubov-de-Gennes equations for a TI-SC heterostructure, where the $\mathbf{k}\cdot \mathbf{p}$-Hamiltonian for Bi$_2$Se$_3$ \cite{Zhang:2009mi, Liu:2010fk} was used to model the TI. For technical reasons, the SC-part of the structure had to be described by a two-band model, with a superconducting conduction band and a split-off valence band, where the conduction band is assumed to be identical to one of the two orbitals of the $\mathbf{k}\cdot \mathbf{p}$-Hamiltonian. The interface is modelled by a step-function dependence of the Hamiltonian parameters. The authors claim that this model would account for the strong-coupling limit of the proximity effect, while the approach of Stanescu et al. was limited to weakly transmitting interfaces. They also reach the conclusion that the Fu-Kane model is a reasonable low-energy approximation, but do not endeavour to extract a surface state Hamiltonian from their calculation. They do, however, perform a self-consistent calculation of the SC-gap and find a suppression in proximity to the interface.

We here propose a different approach to this problem, which can directly be applied to material specific $\bvec{k}\cdot\bvec{p}$- or tight-binding-Hamiltonians obtained from fits to ab-initio band structure calculations. Thus, the starting point is similar to the work of Lababidi et al., but we then proceed by solving the problem with a combination of the finite differences and recursive Green's function methods. This has proven to be convenient tool for transport calculations in low-dimensional conductors \cite{ferry1997transport}. By exploiting that translational invariance is only broken in one direction, we show that it is also a very efficient method for solving the problem discussed here. As opposed to the approach of Stanescu et al., this allows us to treat an infinite system, where potential finite size effects are absent. This is not important if the energy gaps on both sides of the interface are of the order of the band gap, and the interface-states are thus strongly localised. This assumption was made by Stanescu et al. However, a realistic SC-gap is in fact a lot smaller than the band-gap of the TI. The interface-states hence extend into the SC on the scale of the coherence length. In our model, there is also no need to maintain the same Hilbert-space dimensionality on both sides of the interface in the continuum models - which appears to be the reason for the inclusion of  an additional valence band on the SC-side in the work of Lababidi et al. We show that the proximity induced superconducting gap in the surface states can in principle become comparable to the SC-bulk gap, and that the Fu-Kane model is a sensible low-energy theory if the coupling is not too strong.

\section{The model}

We start from the continuum model of the $\mathbf{k}\cdot \mathbf{p}$-Hamiltonian on the TI-side and a Bogoliubov-de-Gennes Hamiltonian with a single parabolic band on the SC-side. We consider an infinite system and the interface is assumed to lie in the $z=0$-plane. We exploit the translational invariance in the $x$-$y$-directions -- which implies the conservation of parallel momentum, $k_{||}$, at the interface -- by Fourier-transforming the Hamiltonians to real-space coordinates only in the $z$-direction ($k_z\mapsto -i\partial_z$). Using the method of finite differences \cite{Datta:1995}, we then map the model onto a 1-dimensional tight-binding Hamiltonian which depends parametrically on $k_{||}$. The coupling between the two parts of the system is modelled by a transfer-Hamiltonian of the form:
\begin{align} \mathcal{H}_{\rm T}=\mathcal{T}\ket{N+1}\bra{N}+\mathcal{T}^\dagger\ket{N}\bra{N+1},\end{align}
coupling the final site $N$ of the TI to the first site $N+1$ of the SC via the matrix element:
\begin{align} \mathcal{T}=\frac{t}{2}\left( \mathcal{P}\mathcal{V}_{\rm TI}+\mathcal{V}_{\rm SC}\mathcal{P}\right). \end{align}

\begin{figure}[t]
\begin{center}
\includegraphics[width=10cm]{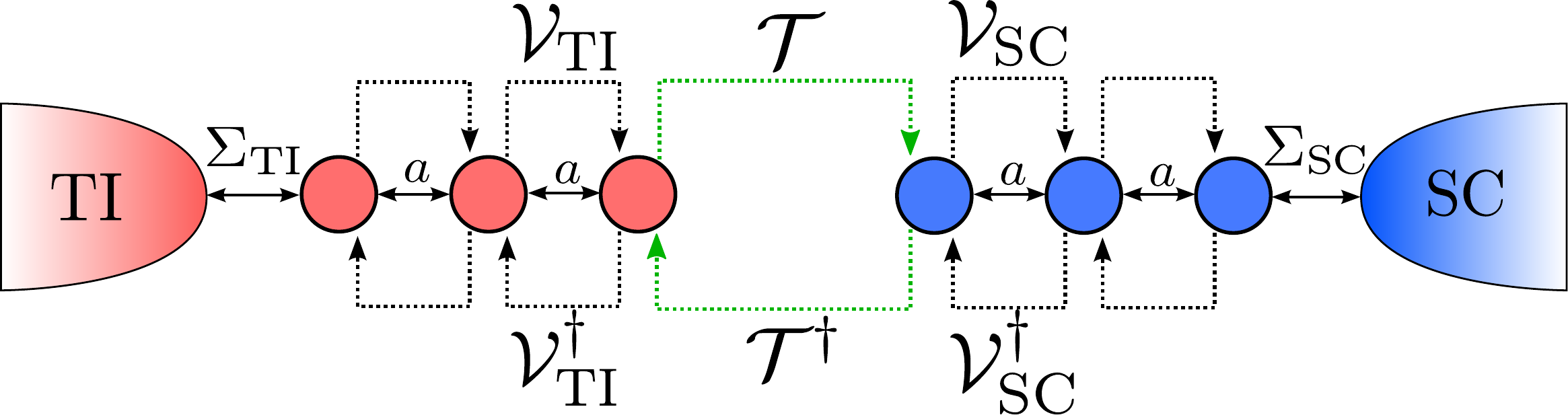}
\end{center}
\caption{\label{fig1} Tight-binding model of the TI-SC heterostructure. A central device, consisting of a one-dimensional tight-binding chain is connected to semi-infinite leads via self-energies that are added to the on-site terms at the respective final sites.}
\end{figure}

Here, $\mathcal{V}_{\rm TI,SC}$ are the respective hopping elements of the TI and the SC derived from the finite differences discretisation (see Fig.~\ref{fig1}). $\mathcal{P}$ is a projector connecting the single band of the SC to the two orbitals of the $\bvec{k}\cdot\bvec{p}$-Hamiltonian. This allows for a greater flexibility in the description of orbital-coupling compared to the approach of Lababidi et al. The inclusion of a second 'dummy' band on the SC side is also not necessary, since we are not forced to maintain the same Hilbert-space dimensionality on both sides of the interface. We found that coupling the SC-band to the TI-orbital with positive parity ('+'-orbital in Ref.~\cite{Zhang:2009mi}) is most effective. We hence consider a $\mathcal{P}$ here, where the matrix elements between the '+'-orbital and the SC-band are equal to 1 and all other elements zero. The parameter $t$ allows us to tune the coupling strength. In the limit $t\rightarrow 0$, the two subsystems are decoupled and the surface state dispersion of a vacuum boundary is recovered. It agrees with the dispersion obtained directly from the continuum model. For $t\rightarrow \infty$, the proximity effect is fully absent as well, but the dispersion is slightly modified compared to the vacuum boundary solution. To investigate the spectral properties of the system, we calculate the local spectral function, which is obtained with a recursive Green's function approach \cite{Metalidis:2005il}. This technique allows us to describe an infinite tight-binding (TB) system by coupling a finite tight-binding chain to semi-infinite leads on both sides of the interface. The devision of the infinite TB-chain into a 'central device' and 'lead' regions is arbitrary, since the procedure of integrating out the leads is exact and not an approximation. The only requirement is that the leads are homogenous, i.e., the interface must be part of the central device (for the case of a homogeneous order order parameter in the SC; otherwise the region of varying order parameter must be included in the central region). The Green's function of the central device coupled to the leads is then formally defined as:
\begin{align} \mathcal{G}(\bvec{k}_{||}, \epsilon)=(\epsilon+i\eta-\mathcal{H}(\bvec{k}_{||})-\Sigma_{\rm SC}(\bvec{k_{||}},\epsilon)-\Sigma_{\rm TI}(\bvec{k_{||}}, \epsilon))^{-1}, \end{align}
where $\Sigma_{\rm SC}$ and $\Sigma_{\rm TI}$ are the self-energies of the SC and TI respectively, and $\mathcal{H}$ is the tight-binding, Bogoliubov-de-Gennes (BdG) Hamiltonian of the central device:
\begin{align} \mathcal{H}(\bvec{k}_{||})=\mathcal{H}_{\rm SC}(\bvec{k}_{||})+\mathcal{H}_{\rm TI}(\bvec{k}_{||})+\mathcal{H}_{\rm T}(\bvec{k}_{||}). \end{align}
The local spectral function is given by:
\begin{align} A_n(\bvec{k}_{||},\epsilon)=-\frac{1}{2\pi a} \mathrm{Im}\ \mathrm{Tr}[\bra{n}\mathcal{G}(\bvec{k}_{||})\ket{n}].\end{align}
$a$ is the lattice spacing of the TB-model. Note that we take the full trace of $\mathcal{G}$, i.e., the Gor'kov Green's function, to map out the whole dispersion of the BdG-Hamiltonian. This implies the presence of hole-like branches in the dispersions we plot. We use $a=0.05\ {\rm \AA}$ and $\eta=10^{-4}\ {\rm eV}$ in the following.

\section{Results}
\begin{figure}
\begin{center}
\includegraphics[width=12cm]{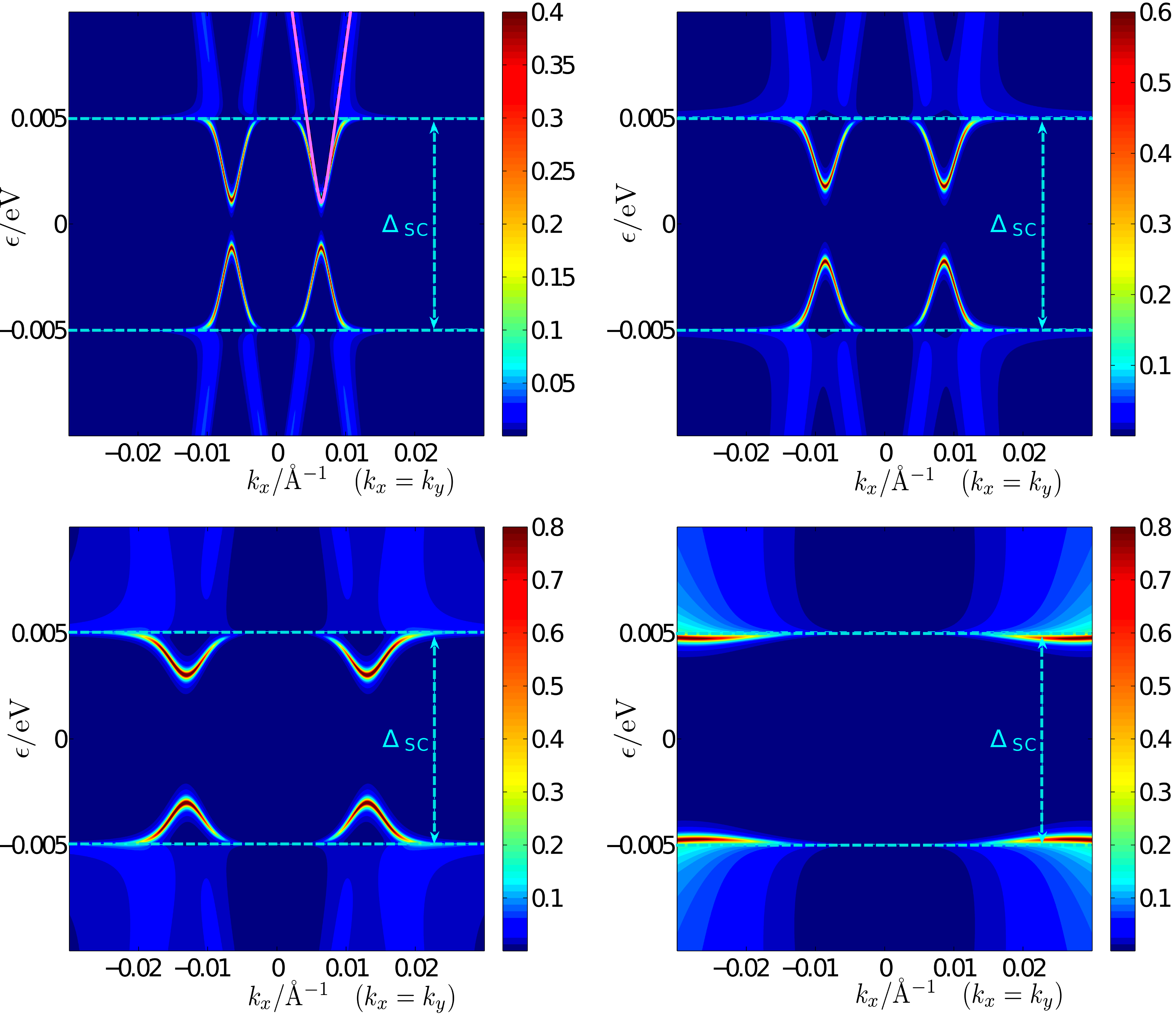}
\end{center}
\caption{\label{fig2} The spectral function $A_N(\bvec{k}_{||},\epsilon)$ as a function of momentum and energy along the cut $k_x=k_y$ for $t=0.92$ (upper-left), $t=0.9$ (upper-right), $t=0.88$ (lower-left) and $t=0.86$ (lower-right). The upper-left plot also shows a tentative fit to one branch of the Fu-Kane dispersion (pink line).}
\end{figure}

We use the parameters provided in Ref.~\cite{Liu:2010fk} for the Bi$_2$Se$_3$ $\bvec{k}\cdot\bvec{p}$-Hamiltonian, but the value of $C_0$ is shifted so that the chemical potential lies at the crossing point of the surface-state dispersion. The chemical potential of Bi$_2$Se$_3$ usually lies above the bottom of the conduction band, i.e., the material is in fact not an insulator \cite{Xia:2009gb, Hsieh:2009mb}. However, chemical doping \cite{Hsieh:2009mb} or gate-control \cite{Chen:2010cq} have been reported to allow for tuning the chemical potential. For the superconductor model, we assume a bulk-gap of $\Delta_{\rm SC}=5\ {\rm meV}$, a Fermi-energy $E_{\rm F}=0.4\ {\rm eV}$, and an effective mass equal to the free electron mass $\hbar^2/2m=3.81\ {\rm eV\AA^2}$. For other values of $m$, we obtain qualitatively the same results.
To analyse the surface-state dispersion, we plot the spectral function $A_N(\epsilon, \bvec{k}_{||})$, $N$ being the site of the TI next to the interface, in Fig.~\ref{fig2} for different values of the interface parameter $t$ and on an energy scale comparable to the SC-gap. The band-gap of the TI is a lot larger than the SC-gap (about 0.3-0.35 eV \cite{Xia:2009gb, Hsieh:2009mb}). Spectral weight from the bulk states of the TI is outside the energy window shown in Fig.~\ref{fig2}. Well-defined interface states in the TI only exist for energies smaller than the SC bulk-gap. Above this energy, the surface states of the TI hybridise with the continuum of itinerant SC-states, which results in the broadening of the spectral function above $\Delta_{\rm SC}$. A tentative fit to one branch of the Fu-Kane model is indicated in the upper-left plot. We observe that the avoided crossing at the chemical potential of the interface states is shifted away from the $\Gamma$-point. This is consistent with the shift of $\mu$ in the Fu-Kane model reported by Lababidi et al. The gap induced in the dispersion of interface states by the superconducting proximity effect is in principle only limited by the SC-gap itself. This strong-coupling limit would, however, require a self-consistent calculation of the SC-gap, which is feasible within our approach, but we have not done it yet. Note also that the dispersion is strongly modified at the energy of the SC-gap and that the Fu-Kane model breaks down in this energy range.
In the range of very strong coupling, which we reach for about $t=0.86-0.84$ with our choice of parameters, the interface states merge into the bulk-DOS of the SC, which leaks into the TI on a length scale given by the band-gap. In this case, there are basically no quasiparticle excitations anymore that can be considered interface states. For even smaller values of $t$ the proximity effect becomes weaker again. 
The calculation of these dispersions is highly effective in our approach, as considering a single site coupled to the semi-infinite leads on the left and right side is sufficient to obtain $A_N(\epsilon,\bvec{k_{||}})$. When a self-consistent calculation of the SC order parameter is included, the TB-chain must be extended into the superconductor until the point where $\Delta_{\rm SC}$ recovers its bulk-value.
\begin{figure}
\begin{center}
\includegraphics[width=12cm]{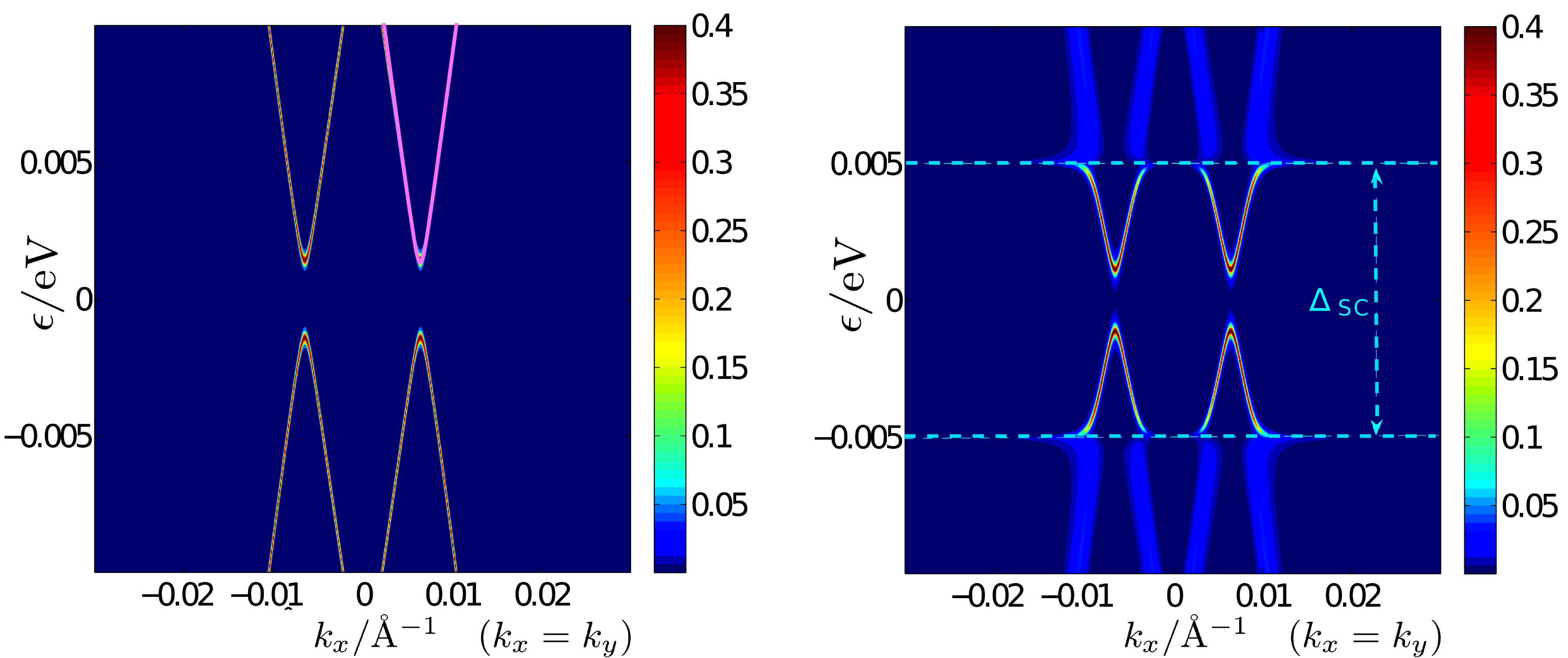}
\end{center}
\caption{\label{fig3} The right plot is the same as the upper-left one in Fig.~\ref{fig2} ($t=0.92$). The left plot shows the same calculation with the assumption$\Sigma_{\rm SC}(\bvec{k}_{||},\epsilon)= \Sigma_{\rm SC}(\bvec{k}_{||}=0,\epsilon=0)$. The indicated fit to the Fu-Kane dispersion matches very well in the energy range shown here and is the same as the one depicted in Fig.~\ref{fig2}.}
\end{figure}

While the agreement of the interface state dispersion is a necessary condition for proving the validity of the Fu-Kane model, it is, in principle, not a sufficient one. To complete this prove, an effective Hamiltonian for the interface states must be extracted from the microscopic calculation. This can be achieved by formulating the proximity of the superconductor in terms of a self-energy, i.e. coupling the final TI-site directly to the SC-lead, and then approximating $\Sigma_{\rm SC}(\bvec{k}_{||},\epsilon)\approx \Sigma_{\rm SC}(\bvec{k}_{||}=0,\epsilon=0)$. This is illustrated in Fig.~\ref{fig3}, were we show that the dispersion thus obtained agrees very well with the Fu-Kane dispersion. The off-diagonal components in particle-hole space of the self-energy $\Sigma_{\rm SC}$ do then indeed have the structure of a singlet, s-wave gap. A surface-state Hamiltonian could now be obtained by considering a \emph{finite} TI tight-binding model, were the self energy $\Sigma_{\rm SC}(\bvec{k}_{||}=0,\epsilon=0)$ is added to the site at the interface. This TB-Hamiltonian can then be diagonalised exactly and projected onto the interface states, which would provide the interface state Hamiltonian.

\begin{figure}
\begin{center}
\includegraphics[width=7.5cm]{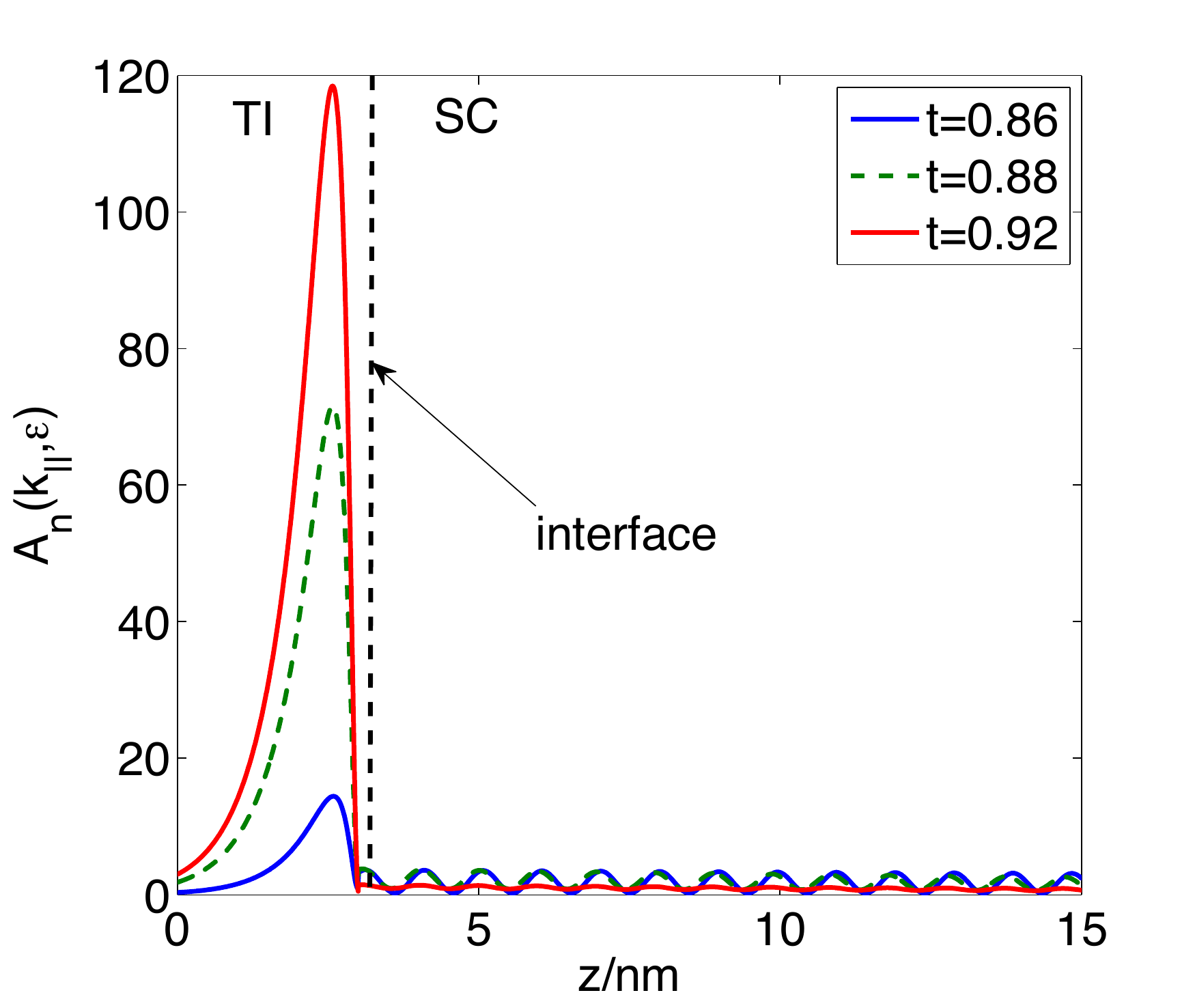}
\end{center}
\caption{\label{fig4} Spatial profile of the states at the bottom of the surface-state dispersion for $t=0.92,0.88,0.86$. Specifically, $k_x=k_y=0.0064\ {\rm \AA^{-1}}$ (red), $0.013\ {\rm \AA^{-1}}$ (green), $0.0275\ {\rm \AA^{-1}}$ (blue); and $\epsilon=1.13\ {\rm meV}$ (red), $=3.03 {\rm meV}$ (green), $=4.74\ {\rm meV}$ (blue). The interface is located at $z=3\ {\rm nm}$.}
\end{figure}

The spatial profile of the states can also be inferred from the spectral function, as shown in Fig.~\ref{fig4}. For weak coupling, the interface states are mainly localised on the TI-side, but as the coupling increases, their weight shifts towards the SC. The decay length on either side of the interface is governed by the energy $\Delta_{\rm SC,TI}-\epsilon$, where $\Delta_{\rm TI,SC}$ is the respective bulk energy gap, and $\epsilon$ is the energy of the state. The band gap $\Delta_{\rm TI}$ is of the order $0.1 {\rm eV}$, while the SC gap is typically two orders of magnitude smaller. Thus, we hardly see any decay on the superconducting side on the interatomic length-scale shown in Fig.~\ref{fig3}, in particular when the coupling is strong and the band bottom of the interface state dispersion is close to the SC-gap.

\section{Conclusions}

We presented an efficient numerical approach to the proximity effect in topological surface states, based on the recursive Green's function technique within the transfer-Hamiltonian approach. The underlying tight-binding model can be derived from ab-initio bandstructure calculations and thus allows for material specific calculations in a straight-forward manner. We showed that the only fundamental limitation to the proximity induced gap in the surface states of the TI is the SC energy-gap itself, but an intermediate coupling regime is most desirable to achieve a superconducting interface state which is well localised and reasonably described by the Fu-Kane model at small energies. We showed that in the case of weak transmission, approximating the self-energy by its zero energy, zero momentum value reproduced the Fu-Kane dispersion to a high precision and argue that this can likely be used to extract an effective interface-Hamiltonian from our calculation.

\section*{Acknowledgements}
RG acknowledges financial support from the Karlsruhe House of Young Scientists (KHYS) and the South East Physics Network (SEPnet).

\bibliography{references_grein_SCES2011}

\providecommand{\newblock}{}
\begin{thebibliography}{10}
\expandafter\ifx\csname url\endcsname\relax
  \def\url#1{{\tt #1}}\fi
\expandafter\ifx\csname urlprefix\endcsname\relax\def\urlprefix{URL }\fi
\providecommand{\eprint}[2][]{\url{#2}}

\bibitem{Koenig:2007hc}
Koenig M, Wiedmann S, Bruene C, Roth A, Buhmann H, Molenkamp L~W, Qi X~L and
  Zhang S~C 2007 {\em Science\/} {\bf 318} 766--70

\bibitem{Hasan:2010fk}
Hasan M~Z and Kane C~L 2010 {\em Rev. Mod. Phys.\/} {\bf 82} 3045--67

\bibitem{RevModPhys.83.1057}
Qi X~L and Zhang S~C 2011 {\em Rev. Mod. Phys.\/} {\bf 83} 1057--110

\bibitem{Fu:2008ly}
Fu L and Kane C~L 2008 {\em Phys. Rev. Lett.\/} {\bf 100} 096407

\bibitem{Nayak:2008tw}
Nayak C, Simon S~H, Stern A, Freedman M and Das~Sarma S 2008 {\em Rev. Mod.
  Phys.\/} {\bf 80} 1083--159

\bibitem{Vorontsov:2008jl}
Vorontsov A~B, Vekhter I and Eschrig M 2008 {\em Phys. Rev. Lett.\/} {\bf 101}
  127003

\bibitem{RevModPhys.36.225}
De~Gennes P~G 1964 {\em Rev. Mod. Phys.\/} {\bf 36} 225--37

\bibitem{Stanescu:2010gb}
Stanescu T~D, Sau J~D, Lutchyn R~M and Das~Sarma S 2010 {\em Physical Review\/}
  B {\bf 81} 241310

\bibitem{Sau:2010cq}
Sau J~D, Lutchyn R~M, Tewari S and Das~Sarma S 2010 {\em Physical Review B\/}
  {\bf 82} 094522

\bibitem{Black-Schaffer:2011fk}
Black-Schaffer A~M 2011 {\em Physical Review B\/} {\bf 83} 060504

\bibitem{Lababidi:2011fk}
Lababidi M and Zhao E 2011 {\em Physical Review\/} B {\bf 83} 184511

\bibitem{Zhang:2009mi}
Zhang H, Liu C~X, Qi X~L, Dai X, Fang Z and Zhang S~C 2009 {\em Nat. Phys.\/}
  {\bf 5} 438--42

\bibitem{Liu:2010fk}
Liu C~X, Qi X~L, Zhang H, Dai X, Fang Z and Zhang S~C 2010 {\em Phys. Rev.\/} B
  {\bf 82} 045122

\bibitem{ferry1997transport}
Ferry D and Goodnick S 1997 {\em Transport in nanostructures\/} (Cambridge
  University Press)

\bibitem{Datta:1995}
Datta S 1995 {\em Electronic Transport in Mesoscopic Systems\/} (Cambridge
  University Press)

\bibitem{Metalidis:2005il}
Metalidis G and Bruno P 2005 {\em Physical Review\/} B {\bf 72} 235304

\bibitem{Xia:2009gb}
Xia Y, Qian D, Hsieh D, Wray L, Pal A, Lin H, Bansil A, Grauer D, Hor Y~S, Cava
  R~J and Hasan M~Z 2009 {\em Nature Physics\/} {\bf 5} 398--402

\bibitem{Hsieh:2009mb}
Hsieh D, Xia Y, Qian D, Wray L, Dil J~H, Meier F, Osterwalder J, Patthey L,
  Checkelsky J~G, Ong N~P, Fedorov A~V, Lin H, Bansil A, Grauer D, Hor Y~S,
  Cava R~J and Hasan M~Z 2009 {\em Nature\/} {\bf 460} 1101--U59

\bibitem{Chen:2010cq}
Chen J, Qin H~J, Yang F, Liu J, Guan T, Qu F~M, Zhang G~H, Shi J~R, Xie X~C,
  Yang C~L, Wu K~H, Li Y~Q and Lu L 2010 {\em Phys. Rev. Lett.\/} {\bf 105}

\end{thebibliography}
\end{document}